# Scientific Machine Learning Framework to Understand Flash Graphene Synthesis

Kianoosh Sattari,[1] Lucas Eddy,[2,3] Jacob L. Beckham,[2] Kevin M. Wyss,[2] Richard Byfield,[1] Long Qian,[2] James M. Tour,[2,4,5*] and Jian Lin[1*]

[1]Department of Mechanical and Aerospace Engineering, University of Missouri, Columbia, Missouri 65211, USA

[2]Department of Chemistry, [3]Applied Physics Program and Smalley-Curl Institute, [4]Department of Materials Science and NanoEngineering, [5]Department of Computer Science and Engineering, NanoCarbon Center and the Welch Institute for Advanced Materials, Rice University, 6100 Main Street, Houston, Texas 77005, United States

*Emails: linjian@missouri.edu; tour@rice.edu

**ABSTRACT**: Flash Joule heating (FJH) is a far-from-equilibrium (FFE) processing method for converting low-value carbon-based materials to flash graphene (FG). Despite its promise in scalability and performance, attempts to explore the reaction mechanism have been limited due to complexity involved in the FFE process. Data-driven machine learning (ML) models effectively account for this complexity, but the model training requires considerable amount of experimental data. To tackle this challenge, we constructed a scientific ML (SML) framework trained by using both direct processing variables and indirect, physics-informed variables to predict the FG yield. The indirect variables include current-derived features (final current, maximum current, and charge density) predicted from the proxy ML models and reaction temperatures simulated from multi-physics modeling. With the combined indirect features, the final ML model achieves an average $R^2$ score of $0.81 \pm 0.05$ and an average RMSE of $12.1\% \pm 2.0\%$ in predicting the FG yield, which is significantly higher than the model trained without them ($R^2$ of $0.73 \pm 0.05$ and an RMSE of $14.3\% \pm 2.0\%$). Feature importance analysis validates the key roles of these indirect features in determining the reaction outcome. These results illustrate the promise of this SML to elucidate FFE material synthesis outcomes, thus paving a new avenue to processing other datasets from the materials systems involving the same or different FFE processes.

# 1. Introduction

Despite the vast applications of graphene, scalable synthesis of graphene remains a tremendous challenge. Among the reported various types of processing methods,[1,2] flash Joule heating (FJH) was introduced in 2020 to synthesize gram-scale graphene from different carbon feedstocks,[3] such as carbon black (CB), metallurgical coke (MC), and waste plastics.[4,5] FJH is an electrothermal process in which Joule heating, driven by capacitors with very high discharge rates, affords gross morphological changes.[2] The generated high temperature (> 3000 K) breaks the chemical bonds and reorganizes the carbon atoms into thermodynamically stable $sp^2$-hybridized graphene sheets.[2] Because the whole process is finished in a sub-second scale, the generated graphene sheets form a metastable state, namely turbostratic graphene, which was termed as flash graphene (FG).[2] Such FG remains highly anisotropic in interlayer arrangements.[3] This feature makes it highly dispersible in solvents and a superior additive for high-performance composites.[3,6]

The scalability of the FJH makes it a promising method for synthesizing the FG, but many unknowns remain in this far-from-equilibrium (FFE) process,[7] making it difficult to establish a processing-property relationship.[8,9] Recently emerged data-driven modeling may provide an alternative solution. In the past several years, some models have been demonstrated to be powerful for tackling a variety of challenges including guiding materials synthesis.[10-14] Furthermore, we recently constructed pure data-driven models to discover the parameters that controlled the FG yield.[15] However, despite reaching an impressive accuracy in predicting the FG yield, the model performance depended on the current parameters measured from the reactions. These intermediate parameters were therefore unavailable as input parameters for prediction if the experiments had not yet been performed. As a result, one cannot apply such models to accurately predict the reaction outcome from a new set of direct input parameters such as voltage, pulse duration, and capacitance prior to experimentation, which makes them impractical for real applications. Thus, developing a ML framework that only uses the direct, controllable experimental parameters to accurately predict reaction outcomes of FJH remains a challenge.

Normally, a data-driven ML model is a "black box", lacking interpretability in mapping the relationship of the input and output. Moreover, model training requires considerable amount of data, a crucial aspect that has been a bottleneck for many materials processing methods such as FJH for FG synthesis.[16-18] In contrast, physics-based models can learn the relationships of the input and output space. Although these models are highly interpretable, they are often difficult to be constructed from complex systems due to a lack of information about the behavior of the system. Thus, the approximations are needed to construct physics-based modeling while they can result in inherent model bias. Therefore, hybrid models which combine data-driven and physics-based modeling can be beneficial in successful model training with limited experimental data while offering high explainability.[19-21] These models can be constructed by modifying the cost functions within data-driven ML models. This modification can adjust the model to obey the outputs of the physics-based models. Daw *et al.* designed a physics-guided neural networks (PGNN) framework that leveraged the output of the physics-based model and observational features by modifying the loss function of the neural network.[22] Raissi *et al.* introduced physics-informed neural networks (PINNs) that obeyed physics laws described by partial differential equations.[23] The additional information gained from the physical laws can train the networks with much less data than needed in pure ML models, thus broadening the applications where data generation is costly.[17] However, in the FJH process, this approach is not practical since there are no defined physical models that can well describe the FFE reaction. Another method of including physics laws into the ML models is to extract physics-informed features from the experiments or theory, which are used as the model input to boost the prediction accuracy.[21, 24] Sun *et al.* synergized the indirect physics-informed descriptors with other direct variables in the ML framework to develop materials with superior properties.[25] To develop thermo-responsive materials, Huang et al. developed a framework where ML models were informed with physicochemical descriptors derived from quantum chemistry calculation.[26] Such physics-based descriptors can serve as the indirect input features to introduce partial physical information to the ML framework.

To better understand the FJH process for FG synthesis, herein, we demonstrate a scientific machine learning (SML) framework that is trained with both direct experimental parameters and indirect physics-informed ones. The goal is to predict the yield of FG. To estimate the reaction temperature from the direct experimental parameters (such as pulse time, voltage, capacitance, and physical information of the input materials), we performed an electrical-thermal multi-physics simulation by COMSOL. Other important indirect features such as the current parameters of final current, maximum current, and charge density were predicted from the proxy ML models. We hypothesize that these current parameters are correlated with the direct experimental parameters and physical properties of the starting materials. To validate this hypothesis, three proxy ML models were trained on these direct parameters to predict those intermediate parameters for a new experiment. In this way, the final ML model does not rely on any intermediate information to predict the reaction outcome if given a new set of direct experimental parameters. Thus, the resulting SML framework is generalizable and needs only limited training samples.[22]

This SML framework has three advantages over our previously reported ML model.[15] First, the models are able to make predictions about the reaction outcome without using any intermediate parameters. This facilitates the use of our prediction model in a model-based optimization algorithm to optimize the FG yield in just a few iterations. Second, the physics-informed descriptors bring additional information to the model, making the black-box ML models more generalizable and accurate in addition to improving the model interpretability. Third, a general methodology of using separate ML models to predict unknown, intermediate reaction parameters from known direct ones is proposed to solve the challenge of lacking enough input features, particularly related to experiments. Thus, such an approach can be readily applied to other materials processed by the same or different methods.

## 2. Results and Discussion

This work used a dataset consisting of 173 separate FJH reactions reported in our previous work.[15] The starting materials were carbon black (CB), metallurgical coke (MC), plastic waste-derived pyrolysis ash (PA), and waste tire-based carbon black (TCB). The structures of the final products were assessed by wide-area Raman mapping. We applied custom-written scripts to analyze the collected Raman spectra (>64 for each FJH reaction), which were used to estimate the FG yield. In the following sections, we first analyze the dataset and explain how to quantify the FG yield. We then elaborate the SML framework. Lastly, we present the model performance in predicting the FG yield.

### 2.1 Analysis of Input and Output Data

Raman spectroscopy has been considered a powerful technique for characterization of carbon structures.[27, 28] Figure 1a shows Raman spectra of amorphous carbon and synthesized FG. The spectrum of amorphous carbon shows two main peaks: D-band at ~1350 cm$^{-1}$ and G-band at ~1600 cm$^{-1}$. The Raman spectrum of FG has a G-peak at ~1580 cm$^{-1}$ and a 2D band at ~2700 cm$^{-1}$. The existence of this 2D band suggests formation of a graphitic lattice.[28] This resonance-enhanced single-Lorentzian 2D band has a narrow full-width at half-maximum (*FWHM*) of ~16 cm$^{-1}$. The $I_{2D}/I_G$ peak intensity ratio reaches up to 17. Both of them suggest good FG crystallinity.[29] From each sample, we collected 100 Raman spectra, which was then averaged to mitigate the variance in the collected individual spectrum. Then, the FG yield can be calculated from these averaged spectra.[15] Figure 1b-e represent the histograms and statistics distribution of the collected samples for each reaction of all the 173 reactions. Specifically, Figure 1b shows the distribution of $I_{2D}/I_G$ with a mean of 0.66 and a standard deviation of 0.17. Figure 1c represents a histogram of average $I_D/I_G$ with a mean of 0.54 and a standard deviation of 0.14. Figure 1d represents the average *FWHM* of the 2D band with a mean of 43.88 cm$^{-1}$ and a standard deviation of 11.55 cm$^{-1}$. Finally, Figure 1e shows a histogram of the FG yield with a mean of 54% and a standard deviation of 27%. Figure S1 shows the yield distribution of the FG synthesized from the four starting materials.

Figure 1f-g show high correlation of $I_{2D}/I_G$ with the FG yield, showing a Pearson's *r* value of 0.73. Figure 1f shows little dependence of the FG yield on $I_D/I_G$, while the value of *FWHM* can well distinguish the samples with a high FG yield (Figure 1g). Most samples have average *FWHM* values of > 40 cm$^{-1}$ and $I_{2D}/I_G$ > 0.75. We also analyzed the FG yield from different starting materials. As illustrated in Figure 1h, the highest FG yield of 72% and the lowest yield of 37% were obtained for CB and MC, respectively. Figure 1i shows the statistical comparison of the FG yield obtained from the four starting materials. Except for MC versus TCB, all other two-way comparisons show significant differences at a set 0.05 significance level.

We hypothesized that the measured parameters including resistant drop, voltage drop, final current, maximum current, charge density, $I_{2D}/I_G$, $I_D/I_G$, *FWHM*, and reaction yield would depend on the starting material. To test the hypothesis, we applied t-distributed stochastic neighbor embedding (t-SNE),[30] a non-linear dimension reduction method, to project all of them in 2D space (Figure 1j). This analysis shows that those obtained from MC and CB are clustered and separated from the others, which indicates that there do exist combination of the parameters for achieving the highest FG yield in CB (Figure 1h). The significant difference in the FG yield from different staring materials indicates that besides the one-hot encoded material type, inclusion of physical information about the starting materials like particle size ($M_{PS}$), resistance ($M_R$), surface area ($M_{SA}$), and percentage of sp$^2$ carbon ($M_{sp2}$) in the input features would greatly increase model accuracy. All these physical properties of the starting materials are tabulated in Table S1.

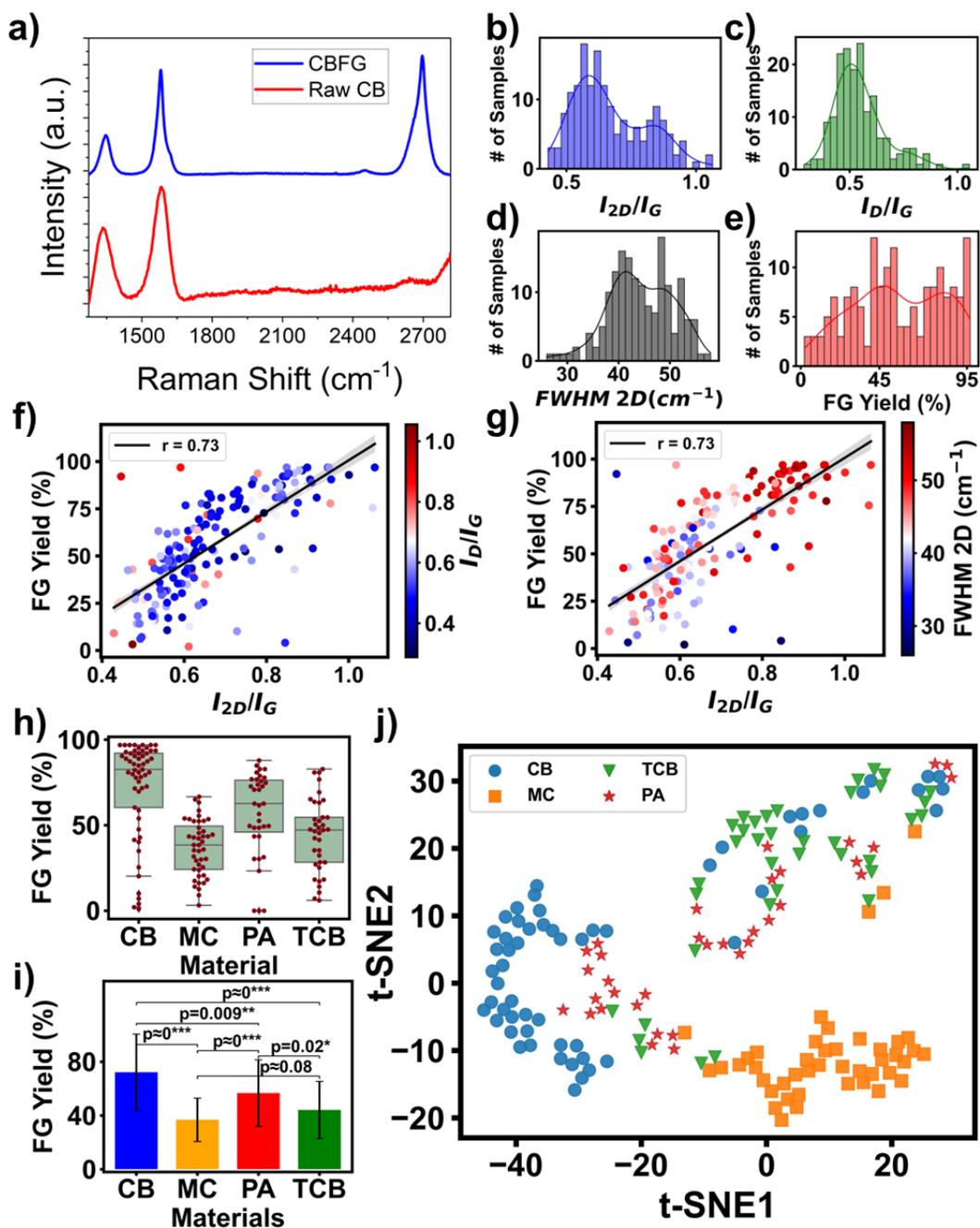

**Figure 1.** (**a**) Raman spectra of flash graphene (FG) synthesized from carbon black and amorphous carbon. (**b-e**) Statistical distribution of $I_{2D}/I_G$ (**b**), $I_D/I_G$ (**c**), *FWHM* of the 2D band (**d**), and FG yield (**e**). Distribution of $I_{2D}/I_G$ versus FG yield in correlation with (**f**) $I_D/I_G$ and (**g**) *FWHM*. (**h**) Distribution of FG yield synthesized from four starting materials. (**i**) Statistical comparison on the mean FG yield from four starting materials. (***), (**), and (*) show significant differences at 0.001, 0.01, and 0.05 levels, respectively. (**j**) t-SNE plots of features in correlation with the four starting materials. The features include resistance drop, voltage drop, maximum current, charge density, $I_{2D}/I_G$, $I_D/I_G$, $FWHM$, and reaction yield.

**2.2 Model Construction and Performance**

The proposed SML framework is shown in Figure 2. The novelty of this framework is that only three types of input features are used for the model development. They include direct reaction parameters such as the properties of starting materials including particle size ($M_{PS}$), resistance ($M_R$), and percentage of sp$^2$ ($M_{sp2}$) and FJH controllable parameters including charge density released from capacitance ($CD_0$), heat ($H$), pulse time ($t$), atmosphere type ($Atm$), and pretreatment voltage ($V_{Pre}$). Using these direct parameters, three proxy models based on XGBoost were trained to predict three intermediate parameters of maximum current normalized by mass ($I_{Max}$), ratio of final current to maximum current ($I_F/I_{Max}$), and charge density ($CD_{IT}$, total charge integrated from the current-time curve and then normalized to mass). In this way, measurement of the time-current curves from a hypothesized experiment is no longer needed. Third, the temperature evolution is simulated from the direct parameters by multi-physics simulation to obtain the maximum temperature ($T_{Sim.}$). Thus, compared to our previous model that predicts the FG yield,[15] more physics-informed input features are used to improve the prediction accuracy and generalizability of the final model. In the following sections, we will elaborate the proxy models, the multi-physics simulation, and the overall architecture of the final prediction model.

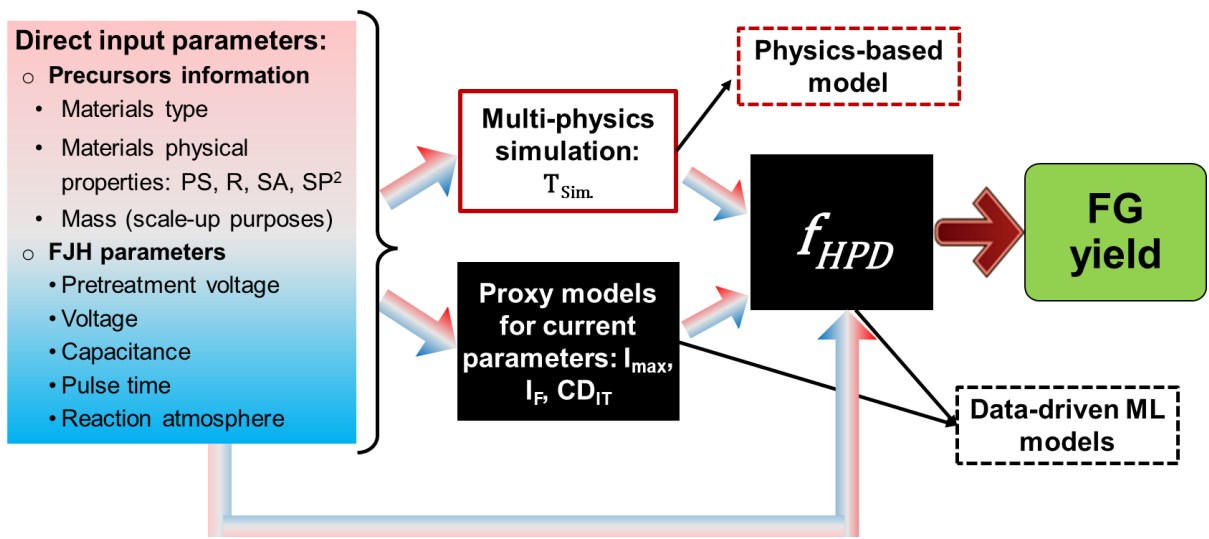

**Figure 2.** Schematic and data flow of the proposed SML framework, where the temperature is simulated by the multi-physics simulation, predicted current parameters, precursor information, and direct FJH parameters are used as the input of the final ML model.

**2.2.1 Proxy models for predicting current parameters.**

The time-current curves are measured from the FJH process. Three parameters of $I_{Max}$, $I_F/I_{Max}$, and $CD_{IT}$ can be extracted from these curves (Figure 3a). The distributions of these current parameters depending on the starting materials were analyzed (Figure S2). Significantly higher $I_{Max}$ values could be realized in the reaction outcomes using MC as the staring material than those in the reactions using other starting materials (Figure S2a). But the higher $I_{Max}$ values do not simply lead to a higher FG yield for the MC samples, as shown in Figure 1h. Figure S3a shows plots of the FG yield vs. $I_{Max}$ grouped by the starting materials. Correspondingly, Pearson's *r* values between $I_{Max}$ and the FG yield for CB, PA, and TCB are 0.41, 0.62, and 0.66, respectively, indicating that they have high correlations, while the correlation of $I_{Max}$ and the FG yield is not significant for MC (Figure S3). The positive correlations between $I_{Max}$ and FG yield for CB, PA, and TCB show that the $I_{Max}$ should pass a threshold value of 1000 (A·g$^{-1}$) for these samples to reach a higher FG yield.

To train the proxy models that predict these three current parameters, the direct reaction parameters, including the properties of starting materials and FJH parameters, serve as the inputs of the models which were trained by a five-fold cross-validation approach. To test the models, 20% of the total samples were used as the never-seen samples. The optimized hyperparameters for these three proxy XGBoost models are listed in Table S2. It is worth mentioning that the inputs to the proxy models can be hypothesized for predicting reaction outcome of a new experiment without performing it. As a result, the trained models can be used to predict the three current parameters for a new reaction. Figure 3b-d shows comparison of the predicted three current parameters from the proxy models versus their true values, from which their Pearson's $r$ values can be calculated to evaluate performance of the proxy models. Pearson's $r$ values of 0.80, 0.78, and 0.77 were obtained for $I_{Max}$, $I_F/I_{Max}$, and $CD_{IT}$, respectively. The high correlations between the predicted and the true values show that the proxy models can predict the output $I_{Max}$, $I_F/I_{Max}$, and $CD_{IT}$ from the direct parameters so that no prior-measurement on the time-current curves for a hypothesized FJH experiment would be needed.

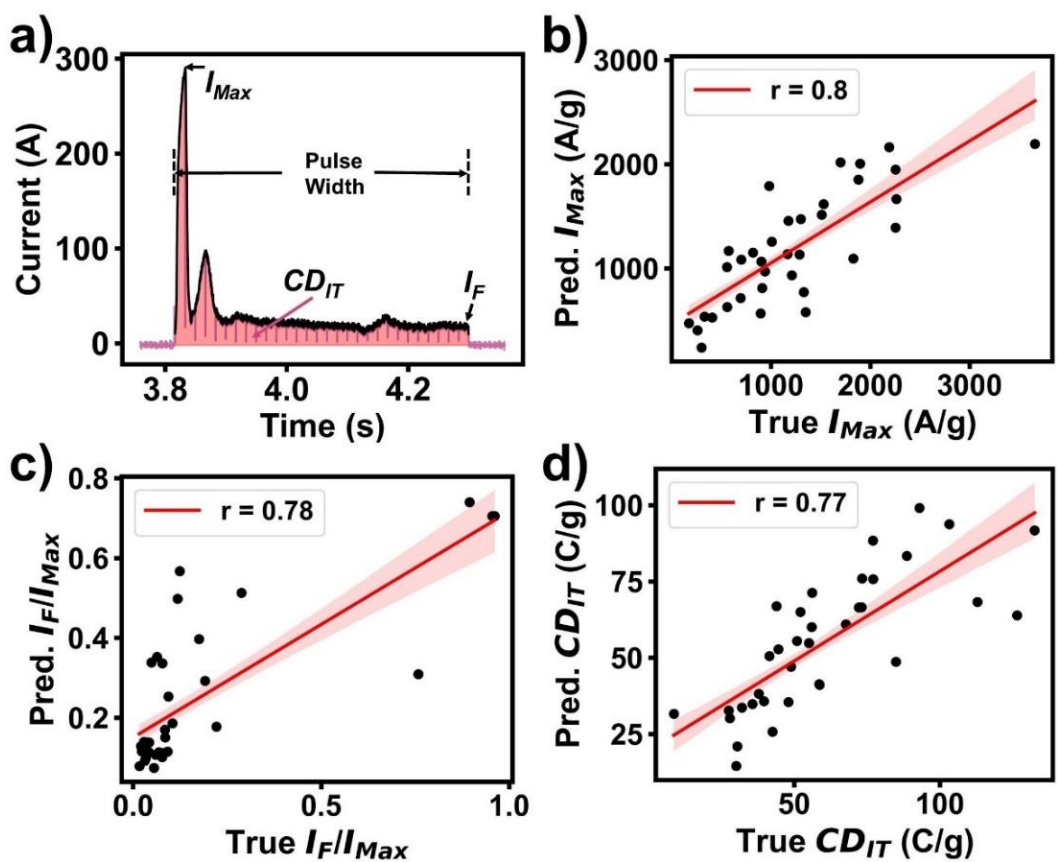

**Figure 3.** (**a**) A represented time-current plot and the current parameters derived from it. Distributions of predicted and true (**b**) $I_{Max}$; (**c**) $I_F/I_{Max}$; and (**d**) $CD_{IT}$ values. Their corresponding Pearson's r values are shown in the figures.

### 2.2.2 Simulation of reaction temperature as a physics-informed input feature

In a FJH process, the electrical energy is rapidly discharged from capacitors, leading to a time-dependent, spatially distributed temperature profile. While temperature is an important parameter that controls the FG yield, we hypothesize that using it as an input feature would improve the predictive accuracy of the model. Deng *et al.* reported the effects of direct reaction parameters like the mass of the starting materials, physical properties of starting materials, pulse time, pulse voltage, pre-treatment voltage, and the maximum temperature achieved in the FJH process.[31] To test the hypothesis, the electrical-thermal multi-physics package in COMSOL was applied to simulate the temperature evolving over the pulse time of each reaction. The maximum temperature of the reaction over the pulse time was then used as an input descriptor, represented as $T_{Sim.}$. In the simulation, the direct input materials and reaction parameters were used. As shown in Figure S4a-b, the FJH quartz tube was simulated as a cylinder with a diameter of 8 mm and a length of ~20 mm. $T_{Sim.}$ over the pulse time for all the 173 reactions are shown in Figure S4c. It shows that the relationship between the temperature and pulse time is not a linear one. There are reactions realizing a higher temperature in a smaller pulse time.

### 2.2.3 Performance of the final model

The predicted current parameters and $T_{Sim.}$ were combined with the direct FJH parameters and precursor information to serve as inputs of six different regression models including linear regression (LR), multilayer perceptron (MLP), Bayesian regression (BR), decision tree (DT), random forest (RF), and eXtreme Gradient Boosting (XGBoost). By using a 5-fold cross-validation method for training and testing, the optimized hyperparameters for these models are listed in Table S3. Figure 4a-b show the coefficient of determination ($R^2$) and root mean squared error (RMSE) for all six tested models in predicting the FG yield. Among them, the XGBoost model reached the highest average $R^2$ score of 0.81 with a standard deviation of 0.05 and the lowest average RMSE of 12.1% with a standard deviation of 2.0% on the testing samples for 5 different train-test splits. Taking a XGBoost model trained from one of the 5 different splits for example, comparison of the predicted FG yields versus the true values was shown in Figure 4c from which an $R^2$ score of 0.84 and RMSE of 11.8% were calculated. As a comparison purposes, we considered a base model that predicts the average value of all testing samples for all the samples. The RMSE for such a naïve model was 29.6% that is significantly higher than that of XGBoost predictions. Samples flashed with CB as the starting material possessed the highest FG yields, while MC-derived FG had the lowest FG yield. Figure 4d shows the relative error (RE) distribution of the predicted FG yields compared with the true values. It shows that 71% of the reactions have the predicted yields of ≤ 10% error of the true values, and only ~11% of the reactions show the predicted FG yields with an error of > 20%. We further examined the distribution of the residuals, a difference of the predicted and the true values. The residuals show a biased toward negative values for samples with the high FG yields, as shown in Figure S5. This indicates that the model usually predicts a lower FG yield value for the reactions resulting in a higher FG yield value, while for the training samples with an average FG yield of 54%, the predictions for unsure testing samples are biased toward the average value.

To test the significance of including the physics-informed features as the input to the model, we trained a separate XGBoost model without using them as the input. As shown in Figure 4e, if the $T_{Sim.}$ is excluded, the $R^2$ score is reduced to 0.79 and RMSE is increased to 13.7% for the same testing dataset. If both the simulated temperature and the predicted current parameters are excluded, the $R^2$ score is greatly decreased to 0.74 and RMSE is increased to 15.1% (Figure 4f). This results because the current parameters may reflect the change of the starting materials' resistance and the contact resistance between the starting materials and the electrode over the pulse time. The temperature is a key parameter that determines the reaction outcome. Consequently, these physics-informed descriptors can offer complementary information to the model with increased the prediction accuracy.

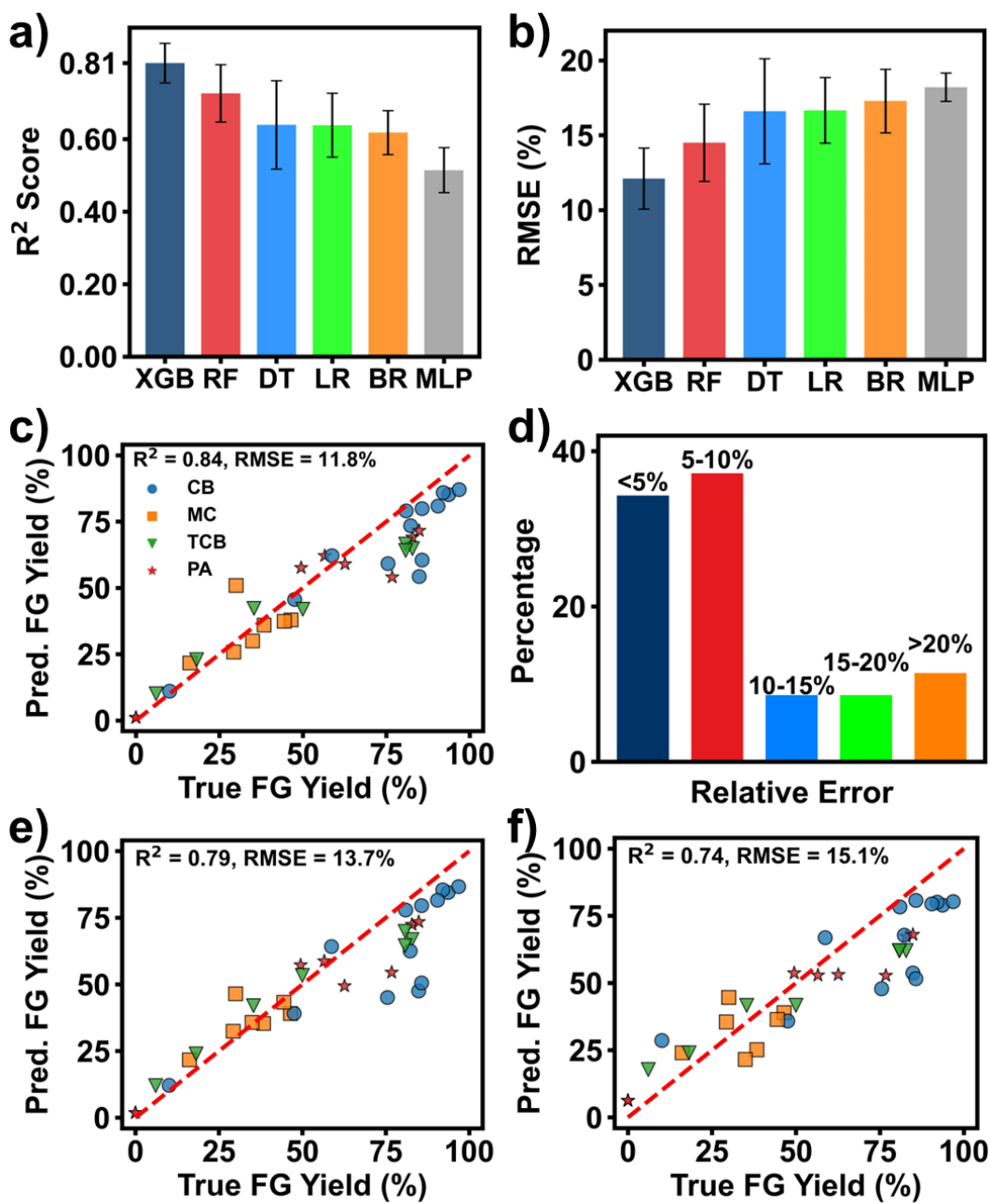

**Figure 4.** Performance of the ML models in predicting the FG yield. (**a**) $R^2$ scores and (**b**) RMSE of the predicted FG yield by the six ML models when using five different train-test splitting ways. The error bars represent the standard deviations from these five testing ways. (**c**) Plot of predicted FG yields by the XGBoost model vs. their true values from different starting materials. (**d**) Relative error distribution of the predicted FG yields shown in **c**. Plot of the predicted FG yields by the XGBoost model vs. their true values after excluding (**e**) $T_{Sim.}$ and (**f**) both $T_{Sim.}$ and predicted current parameters from the direct input parameters.

**2.3 Model Interpretation**

Ranking importance of the input features to the well-trained model in predicting the FG yield would offer additional information about the reaction. The selected features included the $CD_0$, $M_{PS}$, $M_R$, $M_{SP2}$, predicted $I_{Max}$, predicted $I_F/I_{Max}$, predicted $CD_{IT}$, $T_{Sim.}$, $t$, $V_{Pre}$, $Atm$, and $H$. A Pearson's correlation map between these quantitative features is shown in Figure 5a. Low Pearson's $r$ values between any two features indicate that they are quite independent features for the model to afford accurate prediction. For instance, the correlation of the chosen physical properties of the starting materials is low, indicating that they offer complementary information of the materials properties when serving as the input features. In contrast, the surface area has a high Pearson's $r$ value of 0.9 with the particle size, thus we excluded it from the final input features. Figure 5b shows the ranking of the features. $CD_0$ and $T_{Sim.}$ were ranked the Top 2 important features in determining the FG yield, which explains why they play a critical role in the model accuracy (Figure 4). Other features such as the predicted current parameters also have a significant importance in the final prediction. In previous works,[15,32,33] voltage and $CD_0$ were reported to have effects on the transformation rate. Figure 5c shows that the FJH reactions with low $CD_0$ values have a lower FG yield. In contrast, the ones leading to a high FG yield have high $CD_0$ values. This observation agrees well the results shown in these works. In addition, it is found that there is a $CD_0$ threshold value of 100 (C/g) for achieving an FG yield of > 50%. This observation agrees well with other FFE processes. For instance, laser-induced synthesis of graphene from polymers was only initiated when a laser flux reaches a threshold value.[34] Figure 5d shows the importance of $T_{Sim.}$ in predicting the FG yield. It shows that when $T_{Sim.}$ exceeds a threshold value as indicated in green yellow, and red colors, the FG yield is significantly higher than those with low $T_{Sim.}$. A decision tree extracted from the XGBoost model supports the hypothesis that high $T_{Sim.}$ and $CD_0$ are critical in model accuracy for predicting the FG yield (Figure S6). Figure S7 compares $CD_0$ with $C$, $V_0$, and $m$ in correlating with the FG yield. It shows that correlation of FG yield with $CD_0$ is higher than that with $C$, $V_0$, and $m$, which validates the importance of $CD_0$ in the accurate prediction of the FG yield.

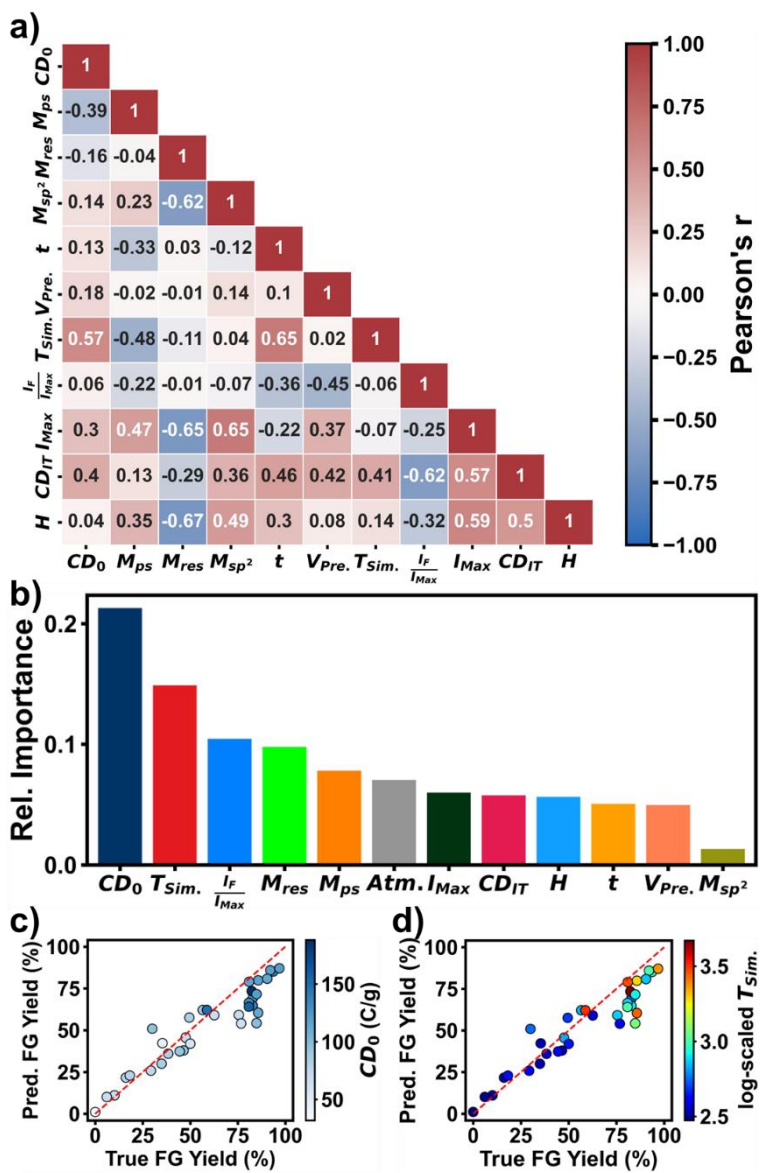

**Figure 5.** Analysis of the input features to the final XGBoost model. (**a**) Quantitative correlation map of the input features. (**b**) Feature importance of the input features. Predicted FG yields versus the true values when correlated with (**c**) $CD_0$ and (**d**) $T_{Sim.}$. In (**d**) $T_{Sim.}$ is in a log scale.

## 3. Conclusion

This study demonstrates a SML framework that bridges a gap between the input processing parameters with the predicted FG yield. Herein, a systematic method of using proxy ML models and multi-physics simulation for extracting physics-informed descriptors, including current-derived properties and simulated temperature, has been developed. These additional input features prove to play a critical role in improving the prediction accuracy of the final ML model. Feature importance analysis further validates this conclusion. Besides the $T_{Sim.}$ and $CD_0$, the selected physical properties of the starting materials are also important features. Explainability of the model by the quantitative analysis offers a glimpse on the reaction mechanism about the FJH. In summary, development of this SML framework offers a methodology of predicting the outcome of new experiments, thus saving the cost and time because of performing unnecessary experiments, which would speed up the FG synthesis. Finally, the methodology can be readily applied to other material systems processed by other processing methods.

## 4. Methods and Experimental Section

*Materials*: Four carbon feedstocks were used as the starting materials. They are carbon black (Cabot BP2000), metallurgical coke (SunCoke Energy Inc., 70–100 mesh size, 150–210 μm grain size), pyrolysis ash (Shangqiu Zhongming EcoFriendly Equipment Co.), and pyrolyzed rubber tire-derived carbon black (Ergon Asphalt and Emulsion Co.). We grinded the materials using a mortar and pestle before and after FJH.

*FJH process*: A custom FJH apparatus was used for all the 173 experiments. Precursor powders with a mass between 100 and 400 mg were sandwiched between two graphite electrodes and compressed inside a quartz tube with an inner diameter of 8 mm. Then, a series circuit with eight 6 mF capacitors (Mouser #80-PEH200YX460BQU2), two 5.6 mF capacitors (80-ALS70A562QH500), and nine 18 mF capacitors (Mouser #80-ALS70A183QS400) were used. Arrangement of capacitors was set to reach the peak capacitance values employed in each flash reaction. To charge the capacitors, the voltage was supplied by a DC source consisting of an AC wall outlet fed through an AC-DC converter. FJH reactions were performed inside a desiccator filled with argon, air, or light vacuum (10 mm Hg) that was used as a categorical descriptor for atmosphere type ($Atm$) among direct input features. After applying the initial voltage, the final voltage was recorded after each reaction. A voltage drop then was calculated by subtracting the final voltage from the initial one. Resistance of the samples were measured before and after each reaction to monitor electrical contact between the electrodes and the samples. Pulse time was modulated by insulated gate bipolar transistors (IGBTs) using programmable millisecond-level delay time. It was connected to a Hall effect sensor through an inductor and controlled via custom LabVIEW scripts. The Hall effect sensor was employed to collect time-current curves. A custom-written Python script was applied on the time-current curves to extract current parameters for the proxy model training.

*Training of machine learning models*: Six different ML models (LR, MLP-R, BR, DT-R, RF-R, and XGB-R) were trained to predict FG yield. The Scikit-Learn package from Python was used for constructing all the models. We kept 20% of the dataset unseen for testing. Cross-validation was applied to optimize the hyperparameters. To test the accuracy of the model for different testing samples, we tried 5 different train/test splits. The results were reported as metrics' mean ± standard deviation.

*Feature engineering*: Twelve selected features included the charge density ($CD_0$) released from the capacitors, starting materials' type ($M$), particle size ($M_{PS}$), resistance ($M_R$), surface area ($M_{SA}$), and percentage of $SP^2$ ($M_{SP2}$), predicted normalized maximum current ($I_{Max}$), predicted ratios of final current to the maximum current ($I_F/I_{Max}$), predicted charge density that is defined as area under the current-time curve normalized by mass ($CD_{IT}$), simulated temperature ($T_{Sim.}$), pulse time ($t$), pre-treatment voltage ($V_{Pre}$), atmosphere type ($Atm$), and nominal heat ($H$) were used as the input features to the final ML models.

$CD_0$, $CD_{IT}$, and $H$ are defined in Eq. 1-3, respectively.

$$CD_0 = \frac{V_0 \times C}{m} \tag{1}$$

$$CD_{IT} = \frac{I \times t}{m} \tag{2}$$

$$H = \frac{V_0^2}{M_R \times t} \tag{3}$$

where $V_0$ is the voltage, $C$ is the capacitance of the capacitors, $m$ is the mass of the starting materials, $M_R$ is the initial resistance of the starting material, and $t$ is the pulse time. $M$ is one-hot encoding for the types of the starting materials. It was only used as input to the proxy models and not in the final model. $CD_{IT}$ was calculated by trapezoidal integration of the time-current curve collected by a Hall effect sensor. Even if $CD_{IT}$ and $CD_0$ have the same units, they include different information about the reaction. $CD_0$ depends on the initial nominal voltage $V_0$, while $CD_{IT}$ conveys information about the voltage drop during the FJH process.

***Evaluation metrics***: The coefficient of determination ($R^2$) is used to evaluate the prediction accuracy of a model as shown in Eq. 4. The Pearson correlation coefficient (r) defined in Eq. 5, on the other hand, measures how the predicted values catch the trend compared to the true values.

$$R^2 = 1 - \frac{\sum_{i=1}^{N}(y_i - \hat{y}_i)^2}{\sum_{i=1}^{N}(y_i - \bar{y})^2} \tag{4}$$

$$r = \frac{\sum_{i=1}^{N}(y_i - \bar{y}) \times (\hat{y}_i - \bar{\hat{y}})}{\sqrt{\sum_{i=1}^{N}(y_i - \bar{y})^2 \sum_{i=1}^{N}(\hat{y}_i - \bar{\hat{y}})^2}} \tag{5}$$

where y is the true values, $\hat{y}$ is the predicted values, $\bar{y}$ is the mean value, and N is the number of samples in both. In Eq. 5, $\bar{\hat{y}}$ is the average of all predicted $\hat{y}$.

Other evaluation metrics including residuals (R), relative error (RE) and root mean squared error (RMSE) are defined in Eq. 6-8, respectively.

$$R = \hat{y} - y \tag{6}$$

$$RE = \frac{|y-\hat{y}|}{y} \times 100\% \tag{7}$$

$$RMSE = \sqrt{\frac{1}{N}\sum_{i=1}^{N}(y_i - \hat{y}_i)^2} \tag{8}$$

where y is the true values, $\hat{y}$ is the predicted values, and N is the number of samples.

*Data inclusion*: At the spectra-level, we included all spectra identified as having a G peak with an SNR of >8 (a maximum in the range of 1500 cm$^{-1}$ < x < 1700 cm$^{-1}$). Spectra not containing a G peak were attributed to poor laser focusing and excluded. For samples to be considered valid, three criteria were checked. First, they should have >64 spectra passing the spectra criterion. Second, they should have a valid recorded current-time curve. Lastly, they should not result in an explosion of the quartz tube.

*FEA simulation on temperature*: The electrical-thermal multi-physics package in COMSOL was applied to simulate the temperature evolving over the pulse time of each reaction. The starting materials mass and particle sizes as well as pulse time, voltage, and capacitance of each reaction were used as the input to the simulation. Also, we considered 140, 130, 120, and 113 (S m$^{-1}$) for the electrical conductivity and 0.4, 1.2, 2.2, and 2.7 (W m$^{-1}$ K$^{-1}$) for the thermal conductivity of the starting materials CB, PA, MC, and TCB, respectively. The applied electrical and thermal conductivity values are in the range of reported experimental values.[35-37] To set up the electrical boundary conditions, one side of the simulated cylinder was considered as ground (0 V) and the other side was applied to the input voltage from dataset. To set up the heat boundary conditions, we considered the room temperature as the initial temperature of the system and applied heat flux at all the edges. After finding the location with the maximum temperature in each reaction, we used the final simulated temperature (end of each pulse time) of the location as the input to the SML as $T_{Sim}$.

**Conflict of Interest**

Universal Matter Inc. has licensed the FG process from Rice University. J.M.T. is a stockholder in that company, but not an officer, director, or employee. Conflicts of interest are managed through regular disclosure to and compliance with the Rice University Office of Sponsored Programs and Research Compliance.

**Associated Content**

The Supporting Information is available free of charge.

Features distribution, simulated temperature of reactions inside the quartz tube, ML models' hyperparameters, starting materials' physical properties (PDF).

**Data and Code Availability**

The used dataset and codes are available at https://github.com/linresearchgroup/SciML_FJH.

**Author Contributions**

J.L. conceived the idea. K.S. designed the framework, constructed the ML models, performed the temperature simulation, and analyzed the data. K.S. wrote the first manuscript which was thoroughly revised by J.L.. R.B. assisted K.S. in model development and manuscript writing. J.L.B. provided discussion on the results. L.E. and K.M.W. performed experiments for data collection. J.M.T supervised L.E, J.L.B., and K.M.W, in the experimental design, data collection as well as revising the manuscript. All authors discussed and commented on the manuscript.

**Acknowledgment**

J. L. and J. M. T. thank U.S. Army Corps of Engineers, ERDC (grant number: W912HZ-21-2-0050) for the financial support. This work was also partially funded by National Science Foundation (award numbers: 1825352 and 2154428), and the Air Force Office of Scientific Research (FA9550-22-1-0526).